\documentclass[structabstract]{aa} 

\usepackage{graphics}
\usepackage{graphicx}
\usepackage{amsmath}
\usepackage{amssymb}
\usepackage{listings}
\usepackage{color}
\usepackage{natbib}
\usepackage{txfonts}
\usepackage[T1]{fontenc}
\usepackage[latin1]{inputenc}

\bibpunct{(}{)}{;}{a}{}{,} 

\author{D. Johansson\inst{1} \and C. Horellou\inst{1} 
\and O. Lopez-Cruz\inst{2} 
\and S. Muller\inst{1} 
\and M. Birkinshaw\inst{3} 
\and J. H. Black\inst{1} 
\and M.N. Bremer\inst{3} 
\and W. F. Wall\inst{2} 
\and F. Bertoldi\inst{4}
\and E. Castillo\inst{2} 
\and H. J. Ibarra-Medel\inst{2} 
}

\institute{Onsala Space Observatory, Department of Earth and Space
  Sciences, Chalmers University of Technology, SE-439 92 Onsala,
  Sweden \and Instituto Nacional de Astrof\'isica, Optica y
  Electr\'onica (INAOE), Tonantzintla, Puebla 72840, Mexico \and HH
  Wills Physics Laboratory, University of Bristol, Tyndall Avenue,
  Bristol BS8 1TL, United Kingdom \and Argelander-Institut f\"ur
  Astronomie, Auf dem H\"ugel 71, D-53121 Bonn, Germany }

\title{Molecular gas and dust in the highly magnified $z\sim
  2.8$ galaxy behind the Bullet Cluster}

\abstract{The gravitational magnification provided by massive
    galaxy clusters makes it possible to probe the physical conditions
    in distant galaxies that are of lower luminosity than those
    in blank fields and likely more representative of the
    bulk of the high-redshift galaxy population.}
{We aim to  constrain the basic properties of molecular gas in a
    strongly magnified submm galaxy located behind the massive Bullet
    Cluster (1E~0657-56). This galaxy (SMM\,J0658) is split into three images, with
    a total magnification factor of almost 100.}  {We used the
  Australia Telescope Compact Array (ATCA) to search for
  $^{12}$CO(1--0)~and $^{12}$CO(3--2)~line emission from SMM\,J0658. We also
  used the SABOCA bolometer camera on the Atacama Pathfinder
  EXperiment (APEX) telescope to measure the continuum emission at
  350~$\mu \mathrm{m}$.}
{CO(1--0)~and CO(3--2)~are detected at {6.8}$\sigma$ and {7.5}$\sigma$
  significance when the spectra toward the two brightest images of the
  galaxy are combined. From the CO(1--0)~luminosity we derive a mass
  of cold molecular gas of ${(1.8 \pm 0.3) \times 10^9} \, M_{\odot}$,
  using the CO to H$_2$ conversion factor commonly used for luminous
  infrared galaxies. This is $45 \pm 25\%$ of the stellar mass.  From
  the width of the CO lines we derive a dynamical mass within the
  CO-emitting region $L$ of ${(1.3 \pm 0.4) \times 10^{10}{ (L/1 {\rm
        kpc}) }} \, M_{\odot}$. We refine the redshift determination
  of SMM\,J0658~ to $z={2.7793 \pm 0.0003}$. The CO(3--2)~to
  CO(1--0)~brightness temperature ratio is ${0.56_{-0.15}^{+0.21}}$,
  which is similar to the values found in other star-forming
  galaxies. Continuum emission at 350~$\mu \mathrm{m}$~from
  SMM\,J0658~was detected with SABOCA at a signal-to-noise ratio of
  3.6. The flux density is consistent with previous measurements at
  the same wavelength by the Herschel satellite and BLAST
  balloon-borne telescope. We study the spectral energy distribution
  of SMM\,J0658~and derive a dust temperature of $33\pm 5$~K and a
  dust mass of ${1.1_{-0.3}^{+0.8}\times 10^{7} \, M_{\odot}}$.}
{SMM\,J0658~is one of the least massive submm galaxies discovered so far.
  As a likely representative of the bulk of the submm galaxy population,
  it is a prime target for future observations.}\date{\today}


\titlerunning{Molecular gas and dust in a highly magnified galaxy at
  $z\sim 2.8$ } \authorrunning{Johansson et al.}

\begin{document}

\maketitle

\section{Introduction}
\label{sec:introduction}

Molecular gas is the raw material from which stars form. Determining
the amount of molecular material and the physical conditions of the
molecular interstellar medium in distant galaxies is important to
understand the cosmic history of star-formation and the evolution of
galaxies (see reviews by \citealt{SolomonVanden-Bout:2005aa} and
\citealt{Walter:2010aa}).  Since the first detection of carbon
monoxide in a high-redshift galaxy twenty years ago (IRAS~F10214+4724
at $z=2.3$; \citealt{BrownVanden-Bout:1991aa, SolomonRadford:1992aa}),
CO and other molecules have been detected in different types of
distant galaxies, indicating the presence of significant reservoirs of
molecular gas: $> 10^8 M_\odot$ in Lyman-break galaxies
(e.g.,~\citealt{StanwayBremer:2008aa},
\citealt{RiechersCarilli:2010aa}), larger amounts (a few times $10^9
M_\odot$) in some submillimeter-faint radio-selected starburst
galaxies (e.g. \citealt{ChapmanNeri:2008aa}), and even larger amounts
(> $10^{10} M_\odot$) in near-infrared--selected star-forming galaxies
(e.g., \citealt{DannerbauerDaddi:2009aa}), submillimeter galaxies
(SMGs) (e.g.,
\citealt{GreveBertoldi:2005aa,DaddiDannerbauer:2009aa,2011MNRAS.412.1913I}),
and in quasars and quasi-stellar objects (e.g.,
\citealt{AlloinKneib:2007aa,CoppinSwinbank:2008aa}). The SMG
population contains extreme objects: high-redshift dust-obscured
star-forming galaxies with rest-frame far-infrared luminosities larger
that 10$^{12}L_\odot$ and star-formation rate of about 1000
$M_\odot$yr$^{-1}$ (e.g. \citealt{SmailIvison:1997aa},
\citealt{BlainSmail:2002aa}).

Whereas the brightest SMGs contribute largely to the star-formation
rate in the Universe, they are not representative of the high-redshift
SMG population as a whole: number counts indicate a steep increase of
the SMG population with decreasing flux density, $S_\nu$: the space
density of SMGs is about $10^4$ deg$^{-2}$ for sources with $S_\nu >
1$~mJy
(e.g. \citealt{HughesSerjeant:1998aa,ChapmanRichards:2001aa,SmailIvison:2002aa}),
and it is about ten times larger for galaxies with $S_\nu> 0.1$~mJy
(e.g. \citealt{Knudsenvan-der-Werf:2008aa}). The abundant sub-mJy
population is difficult to detect { in observations of relatively poor
  angular resolution ($15\arcsec-30\arcsec$) because of confusion
  noise}.  Most detections so far were possible because of
gravitational lensing by a foreground galaxy or a galaxy cluster, that
not only brightens a source but also provides an effective increase in
angular resolution, which lowers the confusion limit
(e.g. \citealt{SmailIvison:1997aa,SmailIvison:2002aa,ChapmanSmail:2002aa,Knudsenvan-der-Werf:2005aa,KnudsenBarnard:2006aa,Knudsenvan-der-Werf:2008aa,JohanssonHorellou:2010ab,WardlowSmail:2010aa,
  RexAde:2009aa,EgamiRex:2010aa,JohanssonSigurdarson:2011aa}).  With
the Atacama Large Millimeter/Submillimeter Array (ALMA), the Large
Millimeter Telescope (LMT) and the Cerro Chajnantor Atacama Telescope
(CCAT), the situation will be transformed, making it possible to
observe faint sources directly. Until then, observations of highly
magnified sources provide a first glimpse into the bulk of the SMG
population.

Few sub-mJy SMGs have been discovered so far: there are seven in
\citet{Knudsenvan-der-Werf:2008aa}'s sample, and five in
\citet{CowieBarger:2002aa}'s sample. \citet{SmailIvison:2002aa}
estimated number counts down to 0.25~mJy, but the low-end counts were
inferred from a Monte Carlo analysis and lower limits on the
magnification of several faint sources. To allow { detection} in a
reasonable amount of observing time, the submm flux density (before
correcting for magnification) should preferrably be larger than
10~mJy, implying a magnification higher than 10. One SMG that
satisfies those criteria is the galaxy at $z\sim 2.5$ lensed by the
massive cluster Abell~2218 (\citealt{Kneibvan-der-Werf:2004aa},
\citealt{KneibNeri:2005aa}). This source has a total magnification of
45 and an intrinsic submm flux of 0.8~mJy.  CO(3--2) was detected in
all three images and CO(7--6) in the brightest one, and a gas mass of
$4.5\times 10^9$ and a star formation rate of about $500 \,
M_{\odot} \, \mathrm{yr}^{-1}$ were inferred \citep{KneibNeri:2005aa}.

Another sub-mJy SMG is the recently discovered galaxy
SMM~J065837.6--555705 (hereafter SMM\,J0658) at $z=2.79$ situated
{near the caustic line of the} Bullet Cluster (1E~0657-56) at $z\sim
0.3$, {and magnified about 100 times}
(\citealt{Bradac:2006fj,2008MNRAS.390.1061W,GonzalezClowe:2009aa},
\citealt{GonzalezPapovich:2010aa} (hereafter G10),
\citealt{JohanssonHorellou:2010ab}). It is the subject of this
paper. The galaxy classifies as a Luminous Infrared Galaxy (LIRG),
with an intrinsic far-infrared luminosity of a few times 10$^{11}
L_\odot$. The source was also detected by BLAST and Herschel; its
infrared spectral energy distribution is consistent with that of a
dusty starburst galaxy \citep{RexAde:2009aa,RexRawle:2010aa}. It is
the brightest source in our APEX LABOCA 870~$\mu \mathrm{m}$~survey of
gravitationally lensed submm galaxies
\citep{JohanssonSigurdarson:2011aa}, in which it was detected with a
significance of $\sim 30\sigma$. When corrected for the gravitational
magnification, it appears to be one of the intrinsically faintest
submm galaxies detected so far ($S_{\mathrm{870 \, {\mu}m}} \sim
0.5$~mJy). The galaxy is strongly lensed, and three images, A through
C, were identified in infrared Spitzer images
(\citealt{GonzalezClowe:2009aa}, G10). Images A and B are separated by
8\arcsec~and have individual magnification factors of $\sim30$ and
$\sim70$ (G10). The third image, C, lies 30\arcsec~away from the
centroid of images A and B, and its magnification factor is smaller
($< 10$), making it too faint for detection in the submm and mm
observations discussed here. Recently, a faint arc extending between
images A and B was discovered in Hubble Space Telescope near-infrared
images (G10). G10 also presented the first spectroscopic redshift of
SMM\,J0658, $z=2.791\pm 0.007$, derived from Polycyclic Aromatic
Carbon (PAH) bands. They also reported the detection of two rotational
lines of H$_2$, and derived a warm molecular gas mass of
$2.2_{-0.8}^{+17} \times 10^8 \; (\mu_{\mathrm{AB}} / 100 )^{-1} \,
M_{\odot}$, where $\mu_{\mathrm{AB}}$ is the total magnification, and
a gas temperature of $377_{-85}^{+68}$ K.

In this paper, we present the first detections of CO(1--0) and
CO(3--2) in SMM\,J0658. The observations were done with the Australia
Telescope Compact Array (ATCA). We use the CO detections to {constrain
  the basic properties of the cold} molecular gas in SMM\,J0658. To
complement the existing Herschel observations we also present APEX
SABOCA observations of the 350~$\mu \mathrm{m}$~continuum that we use to
quantify the dust properties.

Throughout the paper, we adopt the following cosmological parameters:
a Hubble constant $H_0 = 71$~km~s$^{-1}$~Mpc$^{-1}$, a matter density
parameter $\Omega_0 = 0.27$, and a dark energy density parameter
$\Omega_{\Lambda0} = 0.73$. In this cosmology, $z=2.8$ corresponds to
an angular-diameter distance of 1650~Mpc, a luminosity distance of
23800~Mpc and a scale of 8.0~kpc/arcsec\footnote{We used Ned Wright's
  cosmology calculator \citep{Wright:2006aa} available at
  \texttt{http://www.astro.ucla.edu/{\textasciitilde}wright/cosmocalc.html}.}.

\section{Observations and data reduction}
\label{sec:observations}

\subsection{SABOCA (350~$\mu \mathrm{m}$)}
\label{sec:saboca-350micron}

We observed SMM\,J0658~with the Submillimeter APEX Bolometer Camera
(SABOCA), a facility instrument operating in the 350~$\mu
\mathrm{m}$~atmospheric window on the Atacama Pathfinder EXperiment
(APEX\footnote{This publication is based on data acquired with the
  Atacama Pathfinder EXperiment (APEX). APEX is a collaboration
  between the Max-Planck-Institut f\"ur Radioastronomie, the European
  Southern Observatory, and the Onsala Space Observatory.}) telescope
in Chile \citep{Gusten:2006ak}. SABOCA consists of 39 bolometers, has
an instantaneous field-of-view of 1\farcm5, and the main beam has a
full width at half maximum (FWHM) of 7\farcs8
\citep{SiringoKreysa:2010aa}.

SMM\,J0658~was observed in 2010, on June 2nd and 9th and on August 2nd and
5th, as part of observing program O-085.F-9308A. The total observing
time was 7.4 hours, out of which approximately 5.4 hours were spent on
source and the remaining time was used for calibration scans and
pointing. Pointing scans were performed every $\sim 30$ minutes and
the pointing corrections were usually smaller than 2\arcsec. Skydips
(fast scans in elevation at constant azimuthal angle) were performed
at least once during each observing session to estimate the
atmospheric opacity. The weather conditions at the time of the
observations were good, with a median precipitable water vapor level
of 0.46~mm, and standard deviation of 0.05~mm. The derived atmospheric
optical depths were between 0.7 and 1.0. SABOCA was used in scanning
mode to yield a constant noise level within the central arcminute of
the map. In total, 42 scan maps were obtained.

The data were reduced with the \texttt{Crush}
software\footnote{\texttt{Crush} can be used to reduce data from
  several instruments, including the APEX bolometer cameras, and can
  be downloaded from
  \texttt{http://www.submm.caltech.edu/{\textasciitilde}sharc/crush/}}
\citep{Kovacs:2008ab}, following the procedure outlined by Johansson
et al. (2009).

\subsection{ATCA (3~mm and 7~mm)}
\label{sec:interf-mill-observ}

We observed SMM\,J0658~with the Australia Telescope Compact Array
(ATCA\footnote{The Australia Telescope is funded by the Commonwealth
  of Australia for operation as a National Facility managed by
  CSIRO.}) in October and November 2010 and in March 2011 with the aim
to detect the two rotational lines of carbon monoxide
($^{12}\mathrm{C}^{16}\mathrm{O}$) redshifted within the ATCA
frequency coverage: namely, the $J=1\rightarrow 0$ and $J =
3\rightarrow 2$ lines. For gas at $z\sim 2.8$ these lines are
redshifted into the 7~mm and 3~mm bands. The ATCA Compact Array
Broadband Backend (CABB) correlator has two spectral bands with 2 GHz
bandwidth each, which can be arranged in various configurations. The
correlator was set up in the CFB 1M-0.5k mode yielding 2049 channels
each with spectral width 1~MHz. In the 7~mm band, where the
CO(1--0)~line is redshifted to 30.41~GHz, we centered the band on
30.5~GHz, and covered frequencies between 29.5 and 31.5~GHz. In the
3~mm band the CO(3--2)~line is redshifted to 91.23~GHz, and we
centered the first band at 91~GHz, covering the range between 90 and
92~GHz. Because ATCA allows for another band to be observed
simultaneously, we decided to center it at 93.8~GHz to search for the
lines of hydrogen cyanide, HCN, redshifted to 93.53~GHz, and the
formyl ion HCO$^+$, redshifted to 94.13~GHz. The rest frequencies of
the molecular transitions are listed in Table~\ref{tab:lines}.

\begin{table}[t]
  \centering
  \caption{Parameters of the ATCA observations and data.}
  \begin{tabular}[h]{l l}
    \hline
    \hline
    Phase center (J2000): & \\
    Right ascension & 06:58:37.62 \\
    Declination & $-$55:57:04.8 \\
    \hline
    \emph{3 mm band} & \\
    Central observing frequency & 30.5 GHz  \\    
    Configuration & 214H \\
    Bandpass calibrator & 1921--293\\
    Phase calibrator & 0537--441\\
    Primary flux calibrator & Uranus \\
    Primary beam FWHM & 38\arcsec \\
    Synthesized beam\tablefootmark{1} & 2\farcs1 $\times$ 1\farcs7; $82$\degr \\
    Channel velocity width & 3.3~$\mathrm{km} \, \mathrm{s}^{-1}$ \\
    Final velocity resolution & 80~$\mathrm{km} \, \mathrm{s}^{-1}$ \\
    Noise level\tablefootmark{2} & {1.6}~mJy\\
    \hline 
    \emph{7 mm band} & \\
    Central observing frequency & 91 GHz  \\
    Configuration & 750A and 750D \\
    Bandpass calibrator & 0537--441\\
    Phase calibrator & 0724--47\\
    Primary flux calibrator & Uranus \\
    Primary beam FWHM & 110\arcsec \\
    Synthesized beam\tablefootmark{1} & 5\farcs5 $\times$ 1\farcs6; $-8$\degr\\
    Channel velocity width & 9.8~$\mathrm{km} \, \mathrm{s}^{-1}$ \\
    Final velocity resolution & 80~$\mathrm{km} \, \mathrm{s}^{-1}$ \\
    Noise level\tablefootmark{2} & {0.85}~mJy\\
    \hline
  \end{tabular}
  \tablefoot{\tablefootmark{1} Beam full-width at half maximum (FWHM) and position angle, measured
    counter-clockwise from the north. 
    \tablefootmark{2} Measured in
    the final spectra, and referring to a resolution element
    of 80~$\mathrm{km} \, \mathrm{s}^{-1}$.
  }
 \label{tab:obs}
\end{table}

The 7~mm observations were done on November 11-12, 2010. The array was
in the 750A configuration and all six antennas were
available. Additional 7~mm observations were carried out on 2011 March
4 and 7, on Director's Discretionary Time, when the array was in the
750D configuration.  The total observing time was $21$~hours.

The 3~mm observations were done on October 16-17, 2010, in the
H214 array configuration in good weather conditions. On the first
day, antenna 4 was taken offline, so most of the observations were
done with four antennas. The total observing time was 10
hours.

\begin{figure*}[t!]
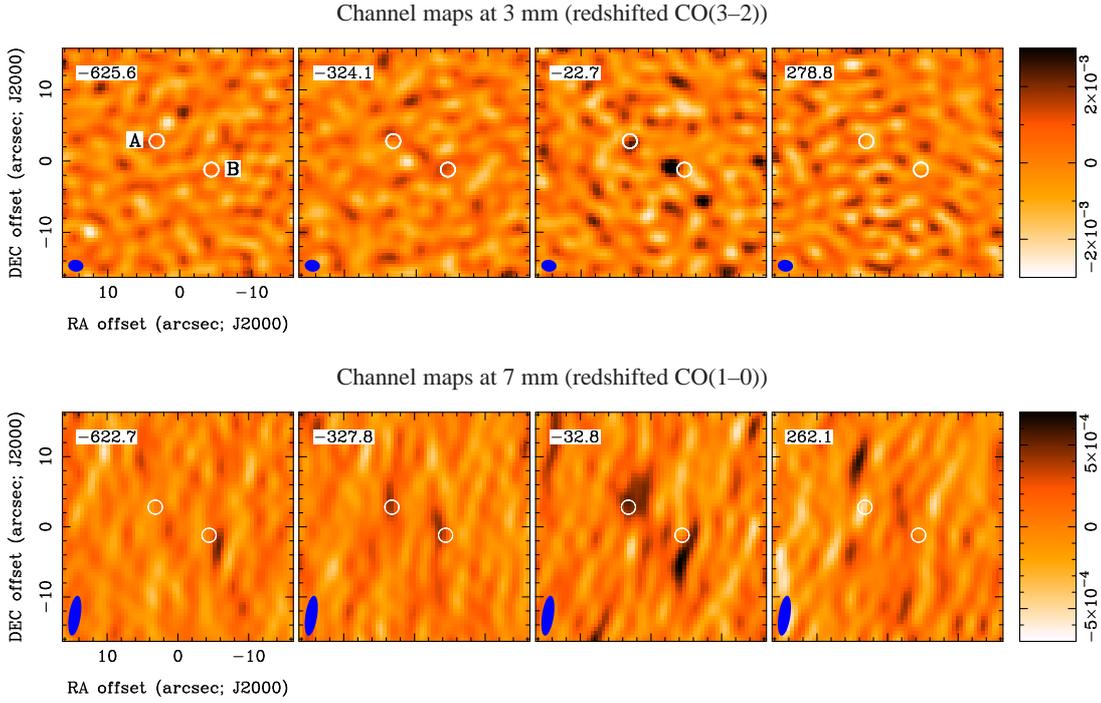

  \centering
  {{Channel maps at 3 mm (redshifted CO(3--2)) }}
  \vspace{3ex}
  \includegraphics[width=3.8cm,angle=-90]{fig2_1.ps}\\
  {{Channel maps at 7 mm (redshifted CO(1--0))}}
  \includegraphics[width=3.8cm,angle=-90]{fig2_2.ps}
  \caption{Channel maps around the CO lines. The white circles
    indicate the positions of the Spitzer images A and B
    \citep{GonzalezClowe:2009aa}. Each channel has a width of $\sim
    300~\mathrm{km} \, \mathrm{s}^{-1}$ and the central velocity of each channel is noted in
    the upper left corners. The beam is shown in the bottom left
    corner and the color bars show the range of surface brightnesses,
    in $\mathrm{Jy\, beam}^{-1}$. \emph{Top row}: Channel maps at
    3~mm, the band into which the CO(3--2)~line is
    redshifted. Emission is clearly seen near image B (the western
    image) in the third panel. Faint emission near image A is seen in
    the third panel. \emph{Bottom row}: Channel maps at 7~mm, the band
    into which the CO(1--0)~line is redshifted. Emission is detected
    close to images A and B and is best seen in the third panel. The
    continuum source discussed in Sect.~\ref{sec:detection-7-mm} lies
    outside the field displayed here. The velocity $v=0$ corresponds
    to a redshift of $z=2.7793$. Positional offsets refer to the phase
    center in Table~\ref{tab:obs}.}
  \label{fig:test}
\end{figure*}

\begin{figure*}[t!]
  \centering
  \includegraphics[width=11cm]{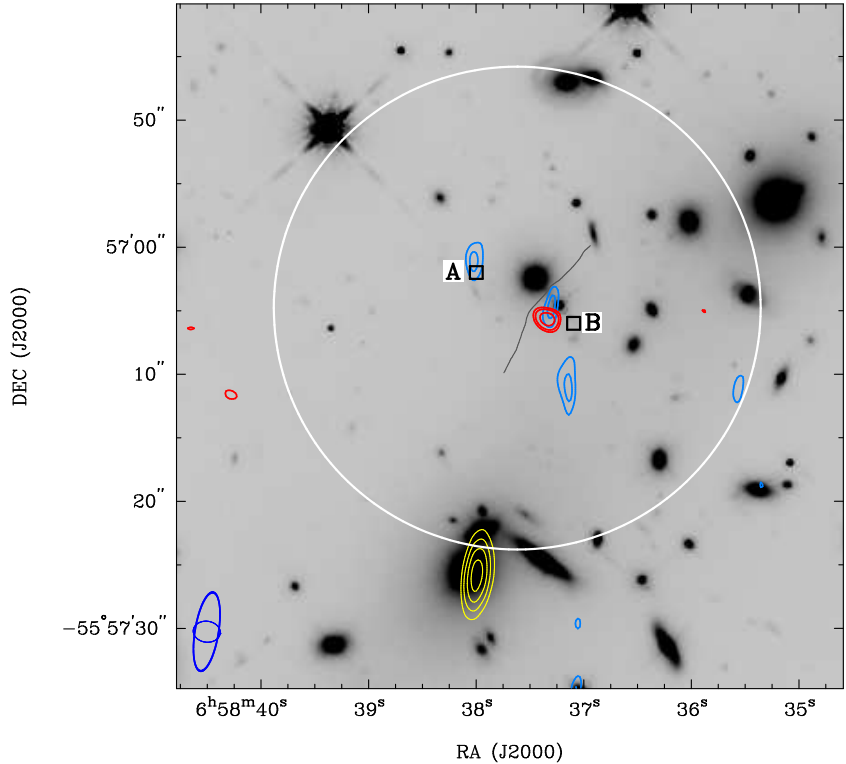}  
  \caption{Hubble Space Telescope WFC3 1.6~$\mu \mathrm{m}$~image of the region
    around SMM\,J0658. The blue contours show the CO(1--0)~integrated
    intensity of both images of SMM\,J0658~and, red contours show the
    CO(3--2)~integrated intensity. The 7~mm continuum emission from
    the $z=0.35$ galaxy to the south is indicated by yellow
    contours. The contour levels of the CO images are 3, 3.5 and
    4$\sigma$, and those of the continuum image are 7, 9, 11 and
    $13\sigma$.  The two black squares indicate the locations of the
    two infrared Spitzer images, A to the east and B to the west
    \citep{GonzalezClowe:2009aa}. CO(1--0)~emission is seen at the
    positions of both images, but CO(3--2)~emission is only seen at
    image B. The low significance of CO(3--2)~at image A can also be
    seen in the spectrum (Fig.~\ref{fig:covel}) The gray line between
    images A and B is the critical line, derived from the lensing
    model for a source redshift of $z=2.7$
    \citep{GonzalezClowe:2009aa}, very similar to the redshift derived
    from the CO observations.  The white circle shows the extent of
    the ATCA FWHM primary beam for the CO(3--2)~observations; the
    primary beam at 7~mm is larger than the field displayed here. The
    offsets seen between the various components are discussed in
    Sect.~\ref{sec:spat-distr-dust}.}
 \label{fig:hstimage}
\end{figure*}

The parameters of the ATCA observations are summarized in
Table~\ref{tab:obs}. The incomplete $uv$-coverage of the 7~mm
observations results in an elongated beam in the north-south direction
(see Table~\ref{tab:obs}, and Figs.~\ref{fig:test} and
\ref{fig:hstimage}).

We used the \texttt{Miriad} software package to reduce the data. We
calibrated each day and frequency band separately, using a standard
calibration procedure. After initial flagging, the bandpass solution
was derived and applied to all sources, followed by gain and flux
calibration. The different calibrators we used are listed in
Table~\ref{tab:obs}. The data for each band were combined to the final
3~mm and 7~mm $uv$-datasets for SMM\,J0658. Images were obtained by
Fourier-transform of these visibility sets, using natural weighting.

At 7~mm we discarded the data from all baselines including the most
remote antenna (located 6~km away from the center of the array)
because of significant phase noise on those baselines. We also flagged
all data during the second half of the first night, when the stability
was poor due to rainfall. For the CLEANing procedure, we selected
regions around the two images of the galaxy and around a bright point
source about 20~arcseconds south of them. This source is discussed in
Sect.~\ref{sec:detection-7-mm}.

At 3~mm we corrected for the opacity of the atmosphere, phase
differences between antennas, and interpolated between the measured
system temperatures. Imaging the 3~mm data gave a map without
noticeable artefacts and no CLEANing of the 3~mm cube was needed.

\section{Results}
\label{sec:anal}

\begin{figure}[t!]
  \centering
  \includegraphics[width=7cm,angle=-90]{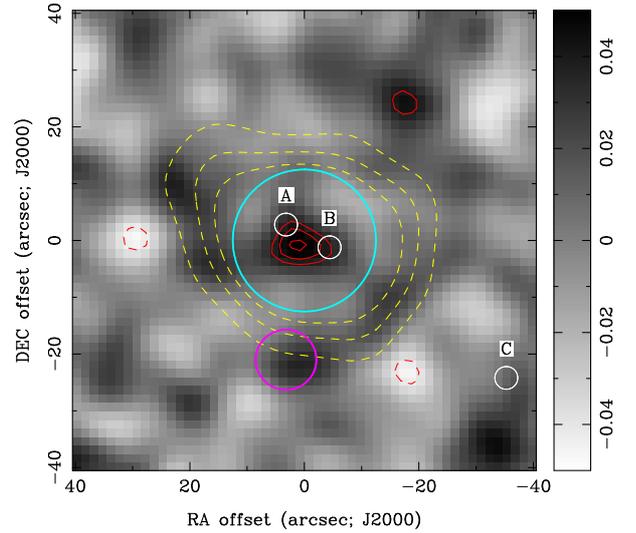}
  \caption{Central region of the SABOCA 350~$\mu \mathrm{m}$~map. The grayscale
    is in Jy/beam. The red contours show significance levels of ${\pm
      2.5, \, 3.0 \;\mathrm{and} \; 3.5\sigma}$ (positive values as
    solid lines and negative values as dashed lines). SMM\,J0658~is
    detected at the center of this image at $3.6\sigma$
    significance. The white circles indicate the positions of the
    three Spitzer images A, B and C \citep{GonzalezClowe:2009aa}. The
    magenta circle has a diameter of 10\farcs6, the FWHM of the SABOCA
    image; it is located at the position of an infrared-bright
    elliptical galaxy at $z=0.35$ \citep{RexRawle:2010aa} which is not
    detected by SABOCA, as expected (the continuum from this galaxy is
    detected in the ATCA 7~mm observations and used to assess the
    astrometry of the CO(1--0) map, see
    Sect.~\ref{sec:detection-7-mm}). The yellow dashed lines
    correspond to the $3, \; 6 \; \mathrm{and} \; 9\sigma$ levels of
    the LABOCA 870~$\mu \mathrm{m}$~detection. The source was detected by
    Herschel \citep{RexRawle:2010aa}: the cyan circle shows the size
    of the Herschel beam at 350~$\mu \mathrm{m}$~(FWHM of 25\arcsec). The
    coordinates of the $(0,0)$-position are the same as those of the
    phase center of the ATCA observations, which are listed in
    Table~\ref{tab:obs}.}
  \label{fig:sablab}
\end{figure}

\begin{table*}[t!]
  \centering
  \caption{Integrated flux densities and upper limits on interesting molecular
    transitions in the ATCA bands.}
  \begin{tabular}[h!]{l c c c c c c}
    \hline
    & & Image A & Image B & {Image A+B} & Image A+B \\
    \hline
    \vspace{1ex}
    Line & $\nu_{\mathrm{rest}}$  & \multicolumn{3}{c}{\ldots \ldots \ldots \ldots \ldots Integrated flux
      \ldots \ldots \ldots \ldots \ldots } &  $L_{\mathrm{line}} \left( \mu_{\mathrm{AB}} / 100
    \right)^{-1}$ & Flux B/A \\
    &  [GHz] & \multicolumn{3}{c}{
      [Jy km s$^{-1}$] } & $[10^8 \,
    \mathrm{K} \, \mathrm{km} \, \mathrm{s}^{-1}~\mathrm{pc}^2 ]$ & \\
    \hline
    $^{12}$CO(1--0) & 115.271 & $0.34\pm0.07$ & $0.29\pm0.07$ &
    $0.63\pm 0.10$ & $22.6 \pm 3.6$ & $0.9_{-0.3}^{+0.5}$ \\
    $^{12}$CO(3--2) & 345.796 & $0.94\pm0.35$ & $2.25\pm0.35$ & $3.18
    \pm 0.50$ & $12.7 \pm 2.0$ & $2.4_{-0.7}^{+1.3}$ \\
    HCN(4--3) & 354.460 & \ldots & \ldots & < {0.7} & < {2.7} & \ldots \\
    HCO$^+$(4--3) & 356.734 & \ldots & \ldots & < {0.7} & < {2.7} &\ldots \\
    CS(7--6) & 342.883 & \ldots & \ldots & < {0.5} & < {2.0} & \ldots \\
    \hline
    \vspace{1ex}
    & & \multicolumn{3}{c}{Brightness temperature ratios ($r_{31}$)} & \\
    \vspace{2ex}    
    & & $0.31_{-0.14}^{+0.22}$ & $0.86_{-0.28}^{+0.45}$ &
    $0.56_{-0.15}^{+0.21}$ & \\
    \hline
 \end{tabular}
 \tablefoot{Uncertainties correspond to the $1\sigma$ level, and 
   upper limits are quoted for $3\sigma$. Integrated flux values are derived
   from fitting point sources at the positions of images A and B in
   the CO(1--0)~and CO(3--2)~maps collapsed for velocities between $-350$
   and $+170$~$\mathrm{km} \, \mathrm{s}^{-1}$, as described in the text. This velocity range was
   determined from the spectral extent of the
   CO(3--2)~spectrum. Measuring the fluxes in the individual and
   combined spectra yields results within 10\% of those
   reported here.} 
 \label{tab:lines}
\end{table*}

\subsection{350 $\mu$m continuum}
\label{sec:saboca-detection}

The 350~$\mu \mathrm{m}$~SABOCA map is shown in Fig.~\ref{fig:sablab}, overlaid
with contours of the signal-to-noise map. The symbols indicate the
location of the infrared counterparts. SMM\,J0658~is detected at 350~$\mu$m
with a peak signal-to-noise ratio of 3.6. The integrated flux density
is $75\pm 21$~mJy, where the uncertainty is the $1\sigma$ noise level
estimated from the data. It does not include the systematic
uncertainty in absolute flux calibration, which is $\sim 20\%$
\citep{SiringoKreysa:2010aa}.

We now compare our measurement to previous detections at
350~$\mu \mathrm{m}$. \citet{RexRawle:2010aa} measured a flux density of
$98.6\pm3.9$~mJy, which does not include calibration uncertainties or
confusion noise, in the Herschel beam of 25\arcsec. The absolute flux
calibration for the SPIRE instrument operating at 350~$\mu \mathrm{m}$~is
better than 15\% \citep{GriffinAbergel:2010aa}. Previously,
\citet{RexAde:2009aa} had measured a flux density at the same
wavelength of $96\pm27$~mJy with BLAST, for which the absolute flux
calibration is better than $10\% - 13\%$. The Herschel, BLAST and
SABOCA measurements are consistent when calibration uncertainties and
confusion noise are taken into account.The SABOCA observations confirm
that the emission comes from SMM\,J0658.

We use the SABOCA and Herschel measurements in the analysis of the
spectral energy distribution (Sect.~\ref{sec:properties-dust-xxx}).

\subsection{CO lines} 

\subsubsection{{ Astrometry of ATCA data}} 
\label{sec:detection-7-mm}

The presence of a bright point-like continuum source in the field of
view of the 7~mm data allows us to assess the astrometry of the 7~mm
map. The source is located at RA 06:58:37.99, Dec --55:57:25.8, which
is 21\arcsec~south of SMM\,J0658, and it is shown in yellow contours in
Fig.~\ref{fig:hstimage}. The elongated 7~mm synthesized beam is shown
in the bottom left corner. When the entire 7~mm dataset is collapsed
in velocity/frequency the signal-to-noise of this source is around
50. The source is located $1.1\arcsec$ southeast of a strong radio
source detected by \citet{Liang:2000hl} in cm-wave ATCA observations
(their source A). Its flux density $S_\nu$ at 7~mm is $1.02 \pm
0.03$~mJy, and together with the cm detections we derive a
centimeter-to-millimeter spectral index $\alpha$ of $-1.02 \pm
  0.05$ ($S_\nu \propto \nu^\alpha$), indicative of a radio
synchrotron emission mechanism.

The extrapolated continuum flux density in the 3~mm band is 0.3~mJy.
This is below the noise level at the edge of the primary beam of the
3~mm map, where the source lies, and it is not detected.

In the HST near-infrared map (see Fig.~\ref{fig:hstimage}), the radio
source has a counterpart in an elliptical galaxy which is a member of
the structure at $z=0.35$ behind the Bullet Cluster
\citep{RexRawle:2010aa}. The near-infrared galaxy is located
  $0.7\arcsec$ northeast of the radio source. The offsets between the
radio source and the $z=0.35$ galaxy in Spitzer maps are also $<
1''$. The offsets to the infrared and infrared counterparts of
the source implies that the astrometry in our 7~mm map is better than
$\sim 1\arcsec$.

In order to remove sidelobe effects from the continuum source, we
subtracted it in the Fourier plane using the Miriad task
\texttt{uvlin}. Imaging this modified dataset produced an improved
CO(1--0)~map of SMM\,J0658.

\subsubsection{{The CO maps}}
\label{sec:position-co-emission}

The CO maps are shown in Fig.~\ref{fig:test} (channel maps) and
Fig.~\ref{fig:hstimage} (contours of CO integrated intensity overlaid
on an HST image in grayscale).

In the channel maps, which are averaged over $\sim300$~$\mathrm{km} \,
\mathrm{s}^{-1}$, emission near the Spitzer images A and B is best
seen in the third panels. In the 3~mm maps, CO(3--2)~is detected at
image~B in the channel centered near 0~$\mathrm{km} \,
\mathrm{s}^{-1}$. A faint peak is seen near image~A. The distance
between the two positions is smaller than that between the Spitzer
images A and B. In the 7~mm maps, CO(1--0)~is detected close to images
A and B in the channel centered near 0~$\mathrm{km} \,
\mathrm{s}^{-1}$~ and also in the one centered near
$-300$~$\mathrm{km} \, \mathrm{s}^{-1}$~, although with lower
significance. The positions of the CO(1--0)~emission are slightly
offset to the north and south of the Spitzer images A and B.

We selected the velocity range from $-350$~$\mathrm{km} \,
\mathrm{s}^{-1}$~to $+170$~$\mathrm{km} \, \mathrm{s}^{-1}$~ to make
images of the integrated CO emission. This is the approximate range
from which the CO(3--2)~emission originates (see
Fig.~\ref{fig:covel}). We measured the integrated intensity of both
transitions and both images in the maps, by fitting the sum of two
point sources convolved with the synthesized beam in each map at the
positions discussed above. Measuring the flux in the extracted spectra
instead yield results which agree within $\sim 10\%$.
CO(1--0)~emission is detected near both Spitzer
images. CO(3--2)~emission, on the other hand, is clearly detected only
near image~B.

Continuum emission at 7~mm is seen from the $z\sim 0.35$ elliptical
galaxy to the south. Offsets between the ATCA and Spitzer counterparts
of SMM\,J0658~are discussed in Sect.~\ref{sec:spat-distr-dust}.

In Table~\ref{tab:lines} we list integrated fluxes and luminosities
for both CO lines, measured from the integrated images. 
The
relation between integrated flux ${F}$ and line luminosity
$L_{\mathrm{line}}$ is
\begin{equation}
  \label{eq:3}
  L_{\mathrm{line}} = 3.25\times 10^{7} \, \left( F/\mu_{\mathrm{AB}}
  \right)  \,
  \nu_{\mathrm{obs}}^{-2} \, D_{\mathrm{L}}^2 (1+z)^{-3} 
\end{equation}
where $\nu_{\mathrm{obs}}$ is the observed (sky) frequency (in GHz) of
the spectral line and $D_L$ is the luminosity distance (in Mpc)
\citep{SolomonVanden-Bout:2005aa}. The dependence on the magnification
value is explicitly included. The CO line luminosities derived from
this equation are expressed in units of K $\mathrm{km} \,
\mathrm{s}^{-1}$~pc$^2$.

\subsubsection{{The CO spectra}}
\label{sec:final-spectra}

\begin{figure}[t!]
  \centering
 \includegraphics[scale = 0.58]{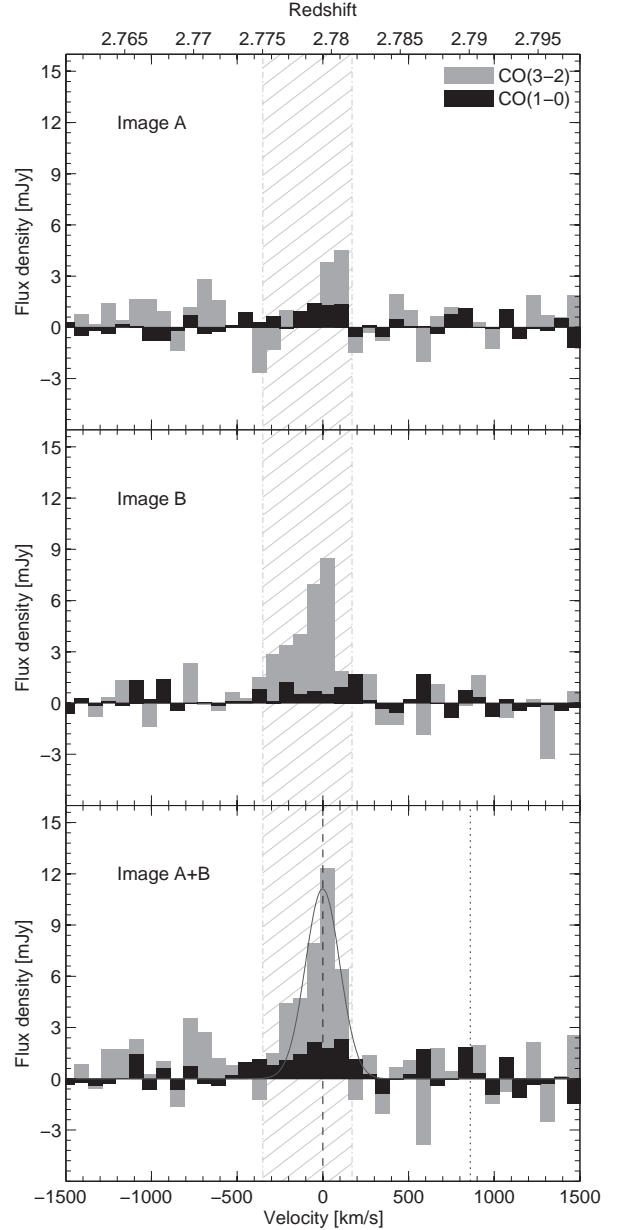}
 \caption{CO(1--0) and CO(3--2) spectra as functions of velocity and
   redshift, showing that both emission lines originate from gas at
   the same systemic velocity. The upper and middle panels show
   individual spectra toward the positions of image A and B. The lower
   panel shows the combined spectrum and the Gaussian model fitted to
   the CO(3--2)~spectrum.  The redshift derived from the CO lines is
   indicated ($z = 2.7793$, dashed line); it differs slightly from the
   redshift derived by G10 from PAH lines ($z\sim 2.79$, dotted
   line). The hashed region shows the spectral extent of the
   integrated images shown as contours in Fig.~\ref{fig:hstimage},
   from which we measured the flux of the CO lines. This velocity
   interval was determined from the combined CO(3--2)~spectrum. The
   velocity resolution of both spectra is 80~$\mathrm{km} \, \mathrm{s}^{-1}$.}
  
  \label{fig:covel}
\end{figure}

The spectra extracted from the positions of images A and B discussed
in the previous section are displayed in Fig.~\ref{fig:covel}. In the
{two upper} panels we show the spectra from the CO(3--2)~and
CO(1--0)~transitions toward both images. Assuming that the spectra
toward images A and B originate from the same galaxy, we can increase
the signal-to-noise ratio of the detections by simply adding the
spectra toward the two images. The total CO(1--0)~and CO(3--2)~spectra
are shown in the lowest panel of Fig.~\ref{fig:covel}.

In the combined spectra (image A + B), CO(3--2)~is detected at
$7.5\sigma$ significance, and CO(1--0)~is detected at $6.8\sigma$
significance.

We used the CO(3--2) spectrum combined for images A and B to determine
the redshift of the galaxy and measure the linewidth by fitting a
single Gaussian. We derived a redshift $z = 2.7793 \pm 0.0003$. This
value is consistent with the CO(1--0)~detection, and slightly lower
than the value measured by G10 (see Fig.~\ref{fig:covel}). This value
is adopted in all $z$-dependent results.

The resulting Gaussian has a velocity width of $\Delta
V_{\mathrm{FWHM}} = 240 \pm 40$~$\mathrm{km} \, \mathrm{s}^{-1}$. The
fit is adequate, with a reduced $\chi^2$ of 1.2 and a probability to
exceed of 20\%. The linewidth is used to estimate the dynamical mass
in Sect.~\ref{sec:molec-dynam-mass}.

A brightness temperature ratio ${r_{31}=0.56_{-0.15}^{+0.21}}$ is
found between the CO(3--2)~and CO(1--0)~lines. This is within the
range of values found in those star-forming galaxies where both CO
lines have been observed. In a sample of 49 local LIRGs,
\citet{LeechIsaak:2010aa} found a median brightness temperature ratio
$r_{31}=0.49$, and ranging between $0.2 < r_{31} < 0.7$.  Few
high-redshift galaxies have been detected in both the CO(1--0)~and
CO(3--2)~line; this is because observation of the CO(1--0)~line in the
$z\sim 2-3$ redshift range has become possible only recently, with the
Expanded Very Large Array (e.g. \citealt{2011MNRAS.412.1913I}), the
CABB correlator of the ATCA, or the Z-receiver at the Green Bank
Telescope. In those SMGs where both CO(1--0)~and CO(3--2)~have been
detected, $r_{31}$-values are similar to that derived for
SMM\,J0658. \cite{SwinbankSmail:2010aa} reported $r_{31}\sim 0.7$ in
the lensed SMG SMMJ2135-0102 at $z\sim 2.3$ and
\cite{HarrisBaker:2010aa} found $r_{31} = 0.68\pm 0.08$ in a sample of
three SMGs at $z\sim 2.5$. \citet{2011MNRAS.412.1913I} found $\langle
r_{31} \rangle = 0.55\pm 0.05$ in four SMGs at $z=2.2-2.5$.
Observations of additional rotational transitions of CO are required
to further constrain the excitation conditions in SMM\,J0658.

\subsection{On the non-detection of dense gas tracers}
\label{sec:nond-dense-gas}

The large bandwidth of the ATCA correlator provides coverage of other
interesting molecular lines (HCN(4--3), HCO$^+$(4--3) and CS(7--6)),
and upper limits on their integrated fluxes and luminosities are given
in Table~\ref{tab:lines}. These molecules have larger critical
densities than CO, and trace denser gas. The flux ratio between
CO(3--2)~and HCO$^+$(4--3) is $>20$ in the Cloverleaf quasar
\citep{WeisHenkel:2003aa,RiechersWalter:2011aa}. In their study of the
lensed SMG SMMJ2135-0102, \citet{DanielsonSwinbank:2010aa} detected
HCN(3--2), and found a flux density ratio of $1/10$ compared to
CO(3--2). Their non-detections of HCO$^+$(4--3) and CS(7--6) also
indicated flux ratios $<1/10$ compared with CO(3--2). This is similar
to our limits on the ratios between CO(3--2)~and HCN, HCO$^+$ and CS,
and thus our non-detections are not surprising.

\section{ Discussion}

\subsection{Dust properties}
\label{sec:properties-dust-xxx}

\begin{figure}[t!]
  \centering
  \includegraphics[width=8cm]{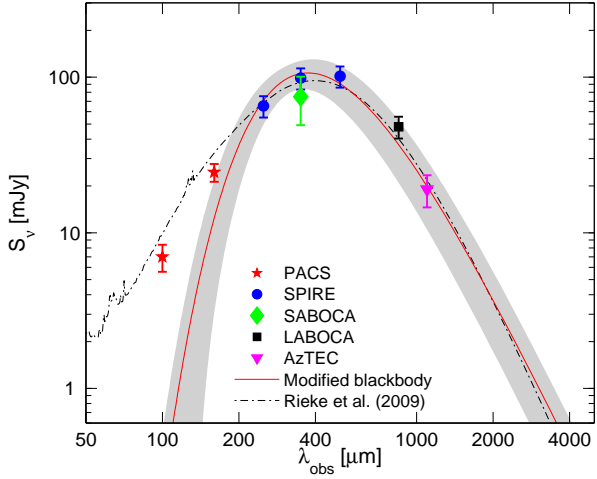}
  \caption{Spectral energy distribution of SMM\,J0658. The solid line
    corresponds to the best-fit model of Eq.~\eqref{eq:2} {to the
      long-wavelength points ($\lambda \ge 250~\mu \mathrm{m}$)}, with
    $\beta = 1.5$, and the shaded area is the 95\% confidence
    interval. The black curve is the LIRG template with far-infrared
    luminosity $2\times 10^{11} \, L_{\odot}$ from
    \citet{RiekeAlonso-Herrero:2009aa}, magnified by a factor
    $\mu_{\mathrm{AB}}= 100$. The data points were taken from
    \citealt{RexRawle:2010aa} (PACS and SPIRE), this work (SABOCA),
    \citealt{JohanssonSigurdarson:2011aa} (LABOCA) and
    \citealt{2008MNRAS.390.1061W} (AzTEC).}
  \label{fig:sed}
\end{figure}

The FIR/submm continuum emission of SMM\,J0658~has been thoroughly studied
and it is established that SMM\,J0658~is a LIRG with a far-infrared
luminosity $L_{\mathrm{FIR}} < 10^{12} \, L_{\odot}$.
\citep{2008MNRAS.390.1061W,RexAde:2009aa,JohanssonHorellou:2010ab,RexRawle:2010aa,Perez-GonzalezEgami:2010aa}.
From the continuum data that include our SABOCA 350~$\mu$m measurement
and the LABOCA 870~$\mu$m measurement of Johansson et al. 2010 we
constructed the FIR/submm spectral energy distribution (SED) displayed
in Fig.~\ref{fig:sed}.

\cite{RexAde:2009aa} used the continuum data measured by BLAST and
AzTEC to estimate a dust temperature in SMM\,J0658~of 32~K. In the
analysis of the more sensitive observations by the SPIRE and PACS
instrument on the Herschel satellite, \cite{RexRawle:2010aa} fitted
SED models but did not derive a dust temperature. To calculate the
dust temperature we therefore fitted a modified blackbody curve to the
SED, following the approach in \citet{RexAde:2009aa}. We adopted the
following functional form of the SED
\begin{equation}
  \label{eq:2}
  S_{\nu} = A \left( \frac{\nu}{\nu_0} \right)^{\beta} B_{\nu} \, (T_{\mathrm{dust}}),
\end{equation}
where $S_{\nu}$ is the flux density, $A$ the amplitude, $\nu$ the
frequency, $\nu_0 = c/250 \, \mu \mathrm{m}$ and $B_{\nu}$ is the
Planck function. We fitted the model to the data using a $\chi^2$
minimization routine, keeping $\beta$ fixed at 1.5. The best-fit curve
is shown in Fig.~\ref{fig:sed} together with the 95\% confidence
interval, and it corresponds to a dust temperature of
$T_{\mathrm{dust}} = 33\pm 5$~K. Changing $\beta$ between 1 and 2
results in changes in the derived dust temperature that are smaller
than the $1\sigma$ confidence interval.

A more elaborate model must be used to fit the full SED, as the PACS
data points fall outside the best-fit curve; however, our aim is not
to accurately model the dust SED, but to derive a lower limit to the
dust temperature. A second hotter dust component within the ISM of
SMM\,J0658~could give rise to an excess power at wavelengths below
100~$\mu \mathrm{m}$~(the detection of hot molecular gas with $T \sim 375$~K
(G10) indicates the possibility of a hot dust component in
SMM\,J0658).  \citet{RexRawle:2010aa} used the LIRG templates of
\citet{RiekeAlonso-Herrero:2009aa} to model the far-infrared/submm SED
in SMM\,J0658. In Fig.~\ref{fig:sed} we show for reference the
\citeauthor{RiekeAlonso-Herrero:2009aa} template for a corresponding
total far-infrared luminosity of $2\times 10^{11} \, L_{\odot}$, magnified
by a factor $\mu_{\mathrm{AB}}= 100$, which fits the observed data
better on the Wien side of the spectrum.

We can also use the \citeauthor{RiekeAlonso-Herrero:2009aa} template
to calculate the estimated continuum flux density for SMM\,J0658~in
the two ATCA bands. The expected flux densities are $< 1$~mJy in the
3~mm band and $<50 \, \mu$Jy in the 7~mm band. This is below our
detection limits and is consistent with our non-detection.

From the SED we also estimate the total mass of cold dust in
SMM\,J0658.  Following \citet{PapadopoulosRottgering:2000aa}, we
calculate the dust mass from
\begin{equation}
  \label{eq:4}
  M_{\mathrm{dust}} = \frac{D_L^2 \, \left( S_{870}/\mu_{\mathrm{AB}} \right)}{(1+z)
    \, k_d (\nu)} \large[ B(\nu,T_{\mathrm{dust}}) - B(\nu,
  T_{\mathrm{CMB}}(z)) \large]^{-1} 
\end{equation}
where $\nu$ is the rest frequency of the observed
870~$\mu \mathrm{m}$~emission, $S_{870}$ is the 870~$\mu \mathrm{m}$~flux density, and
$k_d=0.04 \, (\nu/250 \, \mathrm{GHz})\,^{\beta}$ is the dust
emissivity function. As in the SED fitting, we used $\beta =
1.5$. Together with the dust temperature $T_{\mathrm{dust}}$ derived
above and $T_{\mathrm{CMB}}(z=2.8)=10.3$~K the dust mass is
$M_{\mathrm{dust}} = 1.1_{-0.3}^{+0.8} \times 10^{7} \,
(\mu_{\mathrm{AB}} / 100 )^{-1} \, M_{\odot}$, where the uncertainties are
due to the uncertainty in the dust temperature. The systematic
uncertainty in the normalization and shape of the dust emissivity
function is not included. However, it is considerable and
  likely overshadows the uncertainties in the dust temperature.

\subsection{Molecular and dynamical mass}
\label{sec:molec-dynam-mass}

In this section we estimate the molecular gas mass and the dynamical
mass using the luminosity of the CO lines and their linewidth.

Using CO as a tracer of the more abundant H$_2$~molecule is the main
method to infer the total molecular mass in a molecular cloud or an
entire galaxy \citep*[e.g.][]{DickmanSnell:1986aa}. The relation
between CO luminosity and molecular mass is then $M_{\mathrm{gas}} =
\alpha L_{\mathrm{CO}}$, where the constant of proportionality,
$\alpha$, is $\sim 4.6 \, M_{\odot} \, (\mathrm{K} \,
\mathrm{km} \, \mathrm{s}^{-1}~\mathrm{pc}^2)^{-1}$ for the $J=1-0$ line in Galactic giant
molecular clouds or nearby disk galaxies. For starburst galaxies
(LIRGs and ULIRGs) $\alpha$ is considerably lower ($\alpha \sim 0.8 \,
M_{\odot} \, (\mathrm{K} \, \mathrm{km} \, \mathrm{s}^{-1}~\mathrm{pc}^2)^{-1}$,
\citealt{SolomonDownes:1997aa}). The main reason for the difference is
that in disk galaxies the molecular gas is distributed in individual
virialized molecular clouds, whereas in the centers of starburst
galaxies the CO emission originates from an extended medium and the CO
linewidths are determined by the total dynamical mass in the region
\citep{MaloneyBlack:1988aa,SolomonVanden-Bout:2005aa}.  {For a recent
  discussion of the influence of the physical conditions of the
  interstellar medium on the CO-to-H$_2$ conversion factor in luminous
  IR galaxies see, for instance, \cite{2011arXiv1109.4176P},
  \cite{2012arXiv1202.1803P}}.

Adopting $\alpha = 0.8 \, M_{\odot} \, (\mathrm{K} \,
\mathrm{km} \, \mathrm{s}^{-1}~\mathrm{pc}^2)^{-1}$, and using the luminosity of the
combined CO(1--0)~line displayed in Table~\ref{tab:lines}, we find a
gas mas of $M_{\mathrm{gas}} = {(1.8\pm 0.3) \times 10^9} \, \left(
  \mu_{\mathrm{AB}} / 100 \right)^{-1} \, M_{\odot}$, where the
uncertainties are $1\sigma$ and represent only the statistical
uncertainty in the CO luminosity, not the systematic uncertainty in
the value of $\alpha$, which most likely dominates. Using $\alpha =
3.6\pm 0.8 \, M_{\odot} \, (\mathrm{K} \, \mathrm{km} \, \mathrm{s}^{-1}~\mathrm{pc}^2)^{-1}$
instead, as is appropriate for the Milky Way
(e.g. \citealt{StrongBloemen:1988aa}), nearby spiral galaxies
(e.g. \citealt{MaloneyBlack:1988aa}), and which has also been found in
$z=1.5$ BzK-galaxies \citep{DaddiBournaud:2010aa}, for SMM\,J0658, the
total gass mass is considerably larger, $M_{\mathrm{gas}} = {(8.1\pm
  1.3) \times 10^9} \, \left( \mu_{\mathrm{AB}} / 100 \right)^{-1} \,
M_{\odot}$. Even with the latter value, SMM\,J0658~is a LIRG with low CO
emission and a low gas mass, when compared to intrisically brighter
submm galaxies.

The dynamical mass can be estimated, assuming again that the CO line
originates from an extended medium whose dynamics depend on the total
enclosed mass, from
\begin{equation}
  \label{eq:1}
  M_{\mathrm{dyn}} = {L} \, \Delta V^2 / G,
\end{equation}
(e.g. \citealt{SolomonVanden-Bout:2005aa}), where $L$ is the size of
the region where the CO emission originates from, $\Delta V$ is the
full width at half maximum of the CO line, and G is Newton's constant. 

Due to the moderate signal-to-noise ratio in our data, we refrain from
estimating the size of the CO emitting region. However common values
adopted for SMGs in the literature are $L\sim 1-2$~kpc
(e.g. \citealt{KneibNeri:2005aa,DanielsonSwinbank:2010aa}). Using the
linewidth of the CO(3--2)~line, the dynamical mass is then ${(1.3\pm
  0.4 ) \times 10^{10}} \, \left( {L} / {1\,\mathrm{kpc}} \right) \,
M_{\odot}$.  Uncertainties are due to the uncertainty in the linewidth
and do not include systematic uncertainties, which most likely
dominate. Note that the dynamical mass is to first order independent
of the gravitational magnification.

The properties of SMM\,J0658~are compared with those of other star-forming
galaxies in Sect.~\ref{sec:mass-estimates}.

\subsection{Positional offsets}
\label{sec:spat-distr-dust}

\begin{figure}[t!]
  \centering
  \includegraphics[width=7cm]{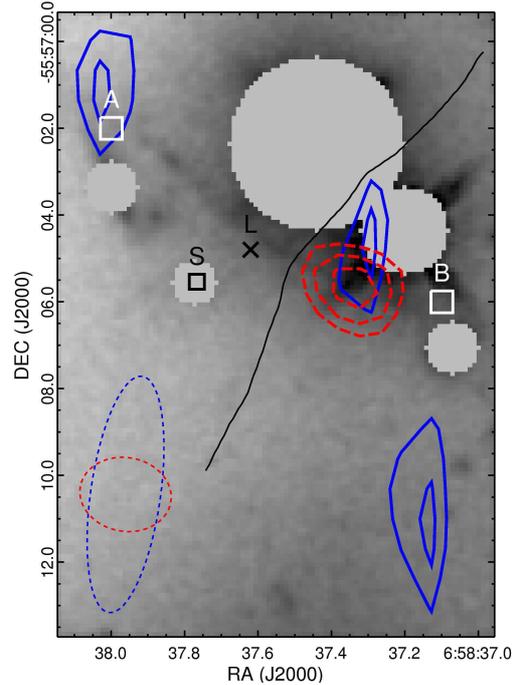}
  \caption{HST image (G10) of SMM\,J0658~overlaid with the two Spitzer
    image positions A and B (white squares). The red and blue contours
    are the same as in Fig.~\ref{fig:hstimage}, and show the
    CO(3--2)~and CO(1--0)~emission. We used GALFIT to subtract a model
    of the elliptical galaxy {at $({\alpha_{\mathrm{2000}},
        \delta_{\mathrm{2000}}})$=(06:58:37.44, $-$55:57:2.4)} and
    masked the region of that galaxy, a nearby star and three other
    objects (gray disks). The faint arc between images A and B is
    visible, roughly orthogonal to the critical line shown in black
    (see Fig.~3 in G10 for a color image). The cross and box markers
    show the centroid of the LABOCA and SABOCA detections. The
    synthesized beams at 3~mm and 7~mm are shown in the bottom left
    corner as red and blue ellipses.}
  \label{fig:arc}
\end{figure}

Offsets of a few arcseconds are seen between the positions of the CO
detections and those of the two Spitzer images of SMM\,J0658, image~A
and image~B.  Those offsets are listed in Table~\ref{tab:pos}, and are
best seen in Fig.~\ref{fig:arc}, which shows the HST WFC3 map of
SMM\,J0658~in grey-scale\footnote{Proposal identifier 11099, principal
  investigator M. Brada{\v c}; (G10).}.  To reveal the faint arc
between the two Spitzer images, we subtracted a model of the
foreground elliptical galaxy using the \texttt{Galfit} code
\citep{PengHo:2010aa}.  We masked out other bright objects near
SMM\,J0658. The positions of the centroids of the SABOCA and LABOCA
detections are shown. Their offset is within the uncertainties of the
observations.\footnote{The positional accuracy when fitting
  two-dimensional Gaussians to sources superimposed on uncorrelated
  Gaussian noise can be written $\Delta x \simeq 0.6 \,
  \mathrm{(S/N)}^{-1} \, \mathrm{FWHM}$
  \citep{Condon:1997aa,IvisonGreve:2007aa}. The accuracy in the fit
  for the SABOCA and LABOCA detections is thus 2\arcsec~and $<
  1\arcsec$ respectively. The pointing accuracy of both instruments is
  2\arcsec~in azimuth and 4\arcsec~in elevation
  \citep{SiringoKreysa:2009aa,SiringoKreysa:2010aa}.}

Contours of the integrated intensity images of CO(3--2)~and CO(1--0)~are
also overlaid. No contour for the CO(3--2)~signal close to image~A is
drawn because of the low significance of the detection.  The CO(3--2)
detection appears about 2$''$ to the east of the Spitzer western image
(image~B).  The CO(1--0) detection, on the other hand, peaks about
5$''$ to the south of image~B.  The CO(1--0) is therefore shifted
relative to the CO(3--2) by about 1.5$''$ in RA and 5$''$ in
declination.  Concerning image~A, a smaller offset of about $2''$ is
seen between the CO(1--0) and the faint CO(3--2) detections. There,
the CO(1--0) seems to be shifted to the north of image~A.

The positional accuracy of our CO(1--0) map has been checked using the
position of the bright radio source 21\arcsec~south of SMM\,J0658, as
described in Sect~\ref{sec:detection-7-mm}; it agrees within one
arcsecond with the position measured by Liang et al. (2000) in cm-wave
ATCA observations and also with the Spitzer and HST
positions.

Clearly, deeper maps are needed to determine whether the observed
offsets in SMM\,J0658~are real or due to a combination of pointing
uncertainties, low signal-to-noise ratio, and the elongated CO(1--0) beam.

\begin{table*}[t!]
  \centering
  \caption{Positions of counterparts of SMM\,J0658.}
  \begin{tabular}[h]{l r r c}
    \hline
    \hline
    & $\Delta \alpha \, ({\arcsec})$ & $\Delta \delta \, (\arcsec)$ & Reference \\
    \hline
    LABOCA~(870~$\mu \mathrm{m}$) & 0.0 & 0.0 & 1 \\
    SABOCA~(350~$\mu \mathrm{m}$)$^{+}$ & $1.3\pm 2.0 \pm 4.5$ & $-0.7\pm2.0 \pm 4.5$ & 2\\
    CO(1--0)~(Image A)$^\star$ & $2.9\pm 0.4$ & $4.0\pm1.1$ & 2\\
    CO(1--0)~(Image B)$^\star$ & $-3.9\pm 0.4$  & $-6.0 \pm 1.2$ & 2\\
    CO(3--2)~(Image A) &$2.9\pm 0.7$ & $2.2\pm 0.5$ & 2\\
    CO(3--2)~(Image B) & $-2.4\pm 0.3$ & $-0.9\pm 0.2$ & 2\\
    IRAC (Image A)$^{\star\star}$ & 3.2 & 2.8 &  3\\
    IRAC (Image B)$^{\star\star}$ & $-$4.4 & $-$1.2 & 3\\
   \hline
  \end{tabular}
  \tablefoot{All angular offsets are stated relative to the LABOCA
    centroid (RA 06:58:37.62; Dec
    $-$55:57:04.8). \\ $^{+}$ The first
    uncertainties come from the Gaussian fit and the second
    uncertainties are the pointing uncertainties. \\ $^{\star}$ We compared the
    astrometry of the CO(1--0)~map to that of the HST image by measuring
    the position of the elliptical galaxy located 21\arcsec~south of
    SMM\,J0658. They agree within 1\arcsec. \\ $^{\star\star}$ The positions
    were taken from \cite{GonzalezClowe:2009aa} who don't quote any
    uncertainties.    \\ References:  1 -- \citet{JohanssonHorellou:2010ab}. 
    2 -- This work. 
    3 -- \citet{GonzalezClowe:2009aa}.}
 \label{tab:pos}
\end{table*}

\subsection{Physical parameters and comparison with other galaxies}
\label{sec:mass-estimates}

Let us compare the properties of SMM\,J0658~with those of other galaxies, in
particular the sub-mJy galaxy at $z\sim 2.5$ (SMM J16359+6612) studied
by \citet{KneibNeri:2005aa} and the brighter lensed galaxy at $z\sim
2.3$ (SMM J2135-0102) studied by \citet{SwinbankSmail:2010aa} and
\citet{DanielsonSwinbank:2010aa}. Those are highly magnified galaxies
with several similarities with SMM\,J0658.

Derived parameters for SMM\,J0658~ and the two other galaxies are
listed in Table~\ref{tab:summ}. The intrinsic submm flux density of
SMM\,J0658~is about 60\% that of the $z\sim 2.5$ galaxy and $\sim 6$
times lower than that of the $z \sim 2.3$ galaxy. The ratios of the
far-infrared luminosities are slightly different: the $z\sim2.5$
galaxy is $\sim 5$ times more luminous, and the $z\sim2.3$ galaxy is
$\sim 8$ times more luminous. As expected, the amount of molecular gas
in SMM\,J0658~is lower than in the two other galaxies. Interestingly,
the star formation efficiency is comparable to that in the ULIRG at
$z\sim 2.3$. The dust mass in SMM J2135-0102 is an order of magnitude
larger than that in the two other galaxies. However, the gas to dust
mass ratio, $M_{\mathrm{gas}}/M_{\mathrm{dust}}$ is $\sim 160$ for
SMM\,J0658, $\sim 240$ for SMM J16359+6612 and $\sim 90$ for SMM
J2135-0102. Given the large systematic uncertainties in determination
of the gas and dust mass, no trends can be identified from the gas to
dust ratios.

\begin{figure}[t!]
  \centering
  \includegraphics[width=8cm]{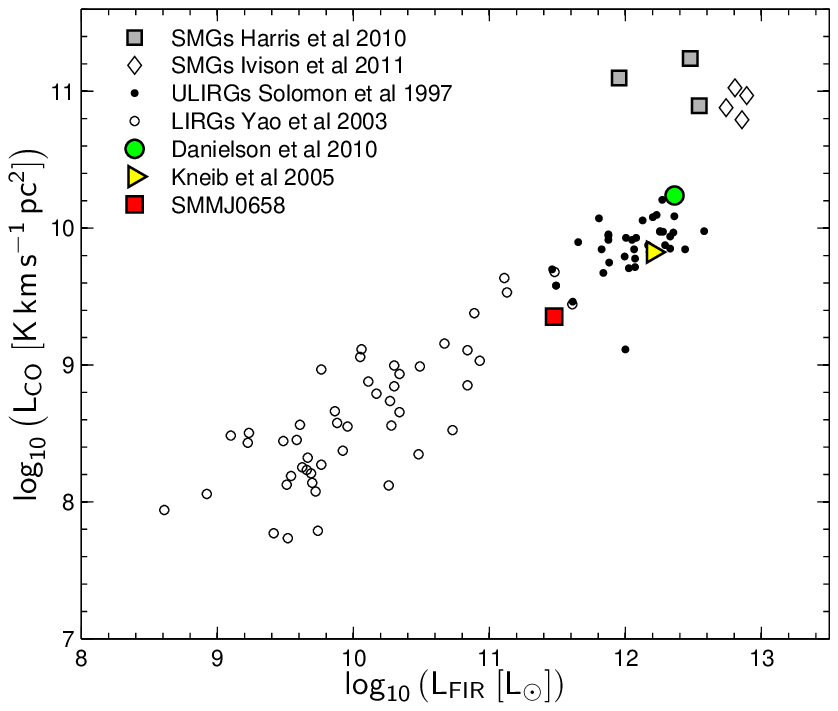}
  \caption{Far-infrared luminosity versus CO(1--0)~luminosity for the
    three galaxies summarized in Table~\ref{tab:summ}:
    SMM\,J0658~(this study), SMM~J16359+6612 \citep{KneibNeri:2005aa}
    and SMM~J2135$-$0102 \citep{DanielsonSwinbank:2010aa}. Local
    {LIRGs and ULIRGs}
    (\citealt{YaoSeaquist:2003aa,SolomonDownes:1997aa}) and
    high-redshift submm galaxies
    \citep{HarrisBaker:2010aa,2011MNRAS.412.1913I} are also
    shown. Because SMM J16359+6612 has not been detected in CO(1--0),
    the CO(3--2)~line luminosity was converted using $r_{31}=0.6$
    (which appears appropriate for SMGs, see
    Sect.~\ref{sec:final-spectra}). SMM\,J0658~is the least
    FIR-luminous high-redshift galaxy.  }
  \label{fig:lfirlco}
\end{figure}

How does SMM\,J0658~compare with larger samples of local and
high-redshift LIRGs and ULIRGs? In Figure~\ref{fig:lfirlco} we show
the CO-line luminosity versus the far-infrared luminosity for the
three galaxies discussed above, and for the nearby luminous infrared
galaxies of \cite{YaoSeaquist:2003aa} and \cite{SolomonDownes:1997aa}
and the bright high-redshift submm galaxies of
\citet{2011MNRAS.412.1913I} and \citet{HarrisBaker:2010aa}
SMM\,J0658~falls on the general correlation and is the least
FIR-luminous high-redshift galaxy in the diagram.

What is the possible duration of its starburst phase in SMM\,J0658?
The star formation rate derived from the far-infrared SED is
100$-$150$\, M_{\odot} \, \mathrm{yr}^{-1} \, [\mu_{\mathrm{AB}} /
100]^{-1}$ (G10). The molecular gas mass divided by the star formation
rate ($M_{\mathrm{gas}}/\mathrm{SFR}$) provides a measure of the
possible duration of the starburst phase, or at least for how long the
starburst phase can last at the current star formation rate and the
current reservoir of molecular gas to form stars from. For
SMM\,J0658~this ratio is 15 to 20~Myr. Note that the starburst
duration is independent of the magnification factor. G10 fitted
stellar population models to the near-infrared SED of SMM\,J0658~and
found that models with a stellar population age $< 90$~Myr or between
1.4~Gyr and 2.6~Gyr give the best fit to the data. They noted however
that the large submm flux density is a strong argument for the lower
age, an argument which is strengthened by the starburst duration
calculated here.

One can estimate the gas mass fraction, $f_{\mathrm{gas}} =
M_{\mathrm{gas}}/(M_{\mathrm{gas}}+M_{\star})$, from the gas mass
derived from the CO lines and the stellar mass derived from
near-infrared measurements. G10 gave $M_{\star} = 4\times 10^9 \,
(\mu_{\mathrm{AB}} / 100 )^{-1} \, M_{\odot}$ and noted that
systematic uncertainties can be a factor of 2 or larger, which makes
them the largest source of uncertainty on the stellar mass. The
influence of systematic uncertainties on derived stellar masses for
near-infrared SEDs is discussed by \citet{ConroyGunn:2009aa}. $M_{\rm
  gas}$ depends of course on the adopted value of the CO-to-H$_2$
conversion factor, $\alpha$. For $\alpha \simeq 0.8 \, M_{\odot} \,
(\mathrm{K} \, \mathrm{km} \, \mathrm{s}^{-1}~\mathrm{pc}^2)^{-1}$ (as
in local ULIRGs), $f_{\rm gas}$ in SMM\,J0658~would be $31\pm12\%$.
For $\alpha \simeq 3.6 \, M_{\odot} \, (\mathrm{K} \, \mathrm{km} \,
\mathrm{s}^{-1}~\mathrm{pc}^2)^{-1}$ (as in the Milky Way or in nearby
spirals, see Sect.~\ref{sec:molec-dynam-mass}), $f_{\mathrm{gas}}
\simeq 67\pm17\%$. Note that the uncertainties on the gas mass
fractions derived here do not include the systematic uncertainty in
the stellar mass determination. The fact that such a large fraction of
the molecular gas has not yet been converted into stars is another
argument in favor of a young system. Such high gas mass fractions are
typical for high-redshift systems: \citet{TacconiGenzel:2010ab} found
$f_{\mathrm{gas}} \simeq 44\%$ in a sample of galaxies around $z\sim
2.3$. Despite the uncertainty in the adopted value of $\alpha$, the
gas mass fraction in SMM\,J0658~is large compared with local galaxies:
for comparison, the gas fraction in a sample of local spiral galaxies
is $\sim 7\%$ \citep{LeroyBolatto:2005aa}.

\begin{table*}[t!]
  \centering
  \caption{Summary and comparison of physical properties of SMM\,J0658~and other highly magnified SMGs.}
  \begin{tabular}[h!]{l c c c}
    \hline
    Source      & SMM~J0658  &    SMM~J16359+6612 &  SMM~J2135$-$0102 \\
    & (1) & (2) & (3) \\
    \hline
    Redshift  & 2.7793  & 2.5174 & 2.3259 \\
    Magnification   & ${80-115}$  & ${45\pm3.5}$  &  $32.5 \pm 4.5$     \\
    Submm flux density (mJy)                                       & $\sim$~0.5  & $\sim$~0.8\tablefootmark{a} & $\sim 3$ \\ 
    $ L_{\mathrm{CO(1-0)}} \; (10^8 \ \mathrm{K \, km \, s^{-1} \,pc^{2}})$ & $22.6\pm 3.6$     & -- & $173 \pm 9$  \\
    $ L_{\mathrm{CO(3-2)}} \; (10^8 \, \mathrm{ K \, km \, s^{-1} \,pc^{2}})$ & $12.7\pm 2.0$  & $37 \pm 2$ & $117.6 \pm 0.9$\\
    $ M_{\mathrm{gas}} \; (10^{9} \, M_{\odot})$ & ${1.8\pm0.3}$ & $4.5 \pm 1.0$ & $14 \pm 1$    \\
    $ M_{\mathrm{dyn}} \; (10^{9} \, M_{\odot})$ & $(13\pm4)(L/1 {\rm kpc})$\tablefootmark{b} & ${15 \pm 3}$ & ${40-80}$ \\
    $ L_{\mathrm{FIR}} \; (10^{12} L_{\odot})$\tablefootmark{c}  &  ${0.3\pm 0.03}$     & $1.6 \pm 0.4$ & $2.3 \pm 0.1$ \\
    $ \mathrm{SFR} \; (M_{\odot} \, \mathrm{yr}^{-1})$   & 100 -- 150  & $\sim 500$ & $400 \pm 20$\\
    $ \mathrm{SFE} \; (L_{\odot} \, M_{\odot}^{-1})$\tablefootmark{d}  & 170  & $\sim 320$ & $165 \pm 7$  \\
    $ M_{\mathrm{dust}} \; (10^{7} \, M_{\odot})$  & $1.1_{-0.3}^{+0.8}$     & $1.9 \pm 0.3$ & $\sim 15$    \\
    $T_{\mathrm{dust}} \; (\mathrm{K})$  & ${33 \pm 5}$  & $51 \pm 3$ & $(30;57 \pm 3)$\tablefootmark{e}\\
    \hline
  \end{tabular}
 \label{tab:summ}
 \tablefoot{All values have been corrected for the individual
   gravitational magnification factors. 
   \tablefoottext{a} Flux density
   measured at 850~$\mu \mathrm{m}$; the other two flux densities were
   measured at 870~$\mu \mathrm{m}$. For a submm spectral index of $\sim 3$ the
   flux difference between the two wavelengths is less than
   3\%. 
   \tablefoottext{b} $L$ is the size of the CO-emitting region. \tablefoottext{c} Rest-frame infrared luminosity between wavelengths
   $\lambda = 8-1000$~$\mu \mathrm{m}$.
   \tablefoottext{d} Star formation efficiency, defined as
   $L_{\mathrm{FIR}}/M_\mathrm{gas}$. \tablefoottext{e} Dust temperatures for
   the extended and clumpy dust component reported by 
   \citet{DanielsonSwinbank:2010aa} for a two-phase
   model. \\
   (1) This work. \\
   (2) \citet{Kneibvan-der-Werf:2004aa,KneibNeri:2005aa}. \\
   (3) \citet{DanielsonSwinbank:2010aa}.}
\end{table*}

\section{Conclusions}
\label{sec:discussion}

This paper presents a study of the molecular gas and dust in one of
the least massive high-redshift galaxies observed so far, the $z=2.8$
submm galaxy SMM\,J0658~lensed by the Bullet Cluster.

\begin{itemize}
\item We detected for the first time rotational transitions of CO from
  SMM\,J0658. The CO(1--0)~to CO(3--2)~brightness temperature ratio is
  comparable to that observed in other high-redshift star-forming
  galaxies.
\item We revise the redshift estimated by
  \citet{GonzalezPapovich:2010aa} from $z=2.791\pm 0.007$ to
  $z=2.7793\pm 0.0003$.
\item The mass of cold molecular gas is estimated to be between $1/3$
  and $2/3$ of the total baryonic mass.
\item From the linewidths of the CO(3--2)~transition we derive a
  dynamical mass of $(1.3 \pm 0.4)\times 10^{10} \, M_{\odot}$, {for a
    CO-emitting disk with a physical size of 1 kpc}.
\item The derived molecular and dynamical masses are consistent with a
  galaxy less massive than the Milky Way. SMM\,J0658~is also less
  massive than most star-forming galaxies observed at high redshift so
  far
  (e.g. \citealt{TacconiGenzel:2010ab,GreveBertoldi:2005aa,SwinbankSmail:2010aa}).
\item Using ground-based 350~$\mu \mathrm{m}$~imaging we detected
  continuum radiation from SMM\,J0658. The signal-to-noise ratio is
  only 3.6, but {the detection} makes it possible to confirm that the
  emission previously seen in the larger beams of BLAST and Herschel
  (\citealt{RexAde:2009aa}, \citealt{RexRawle:2010aa}) comes from
  SMM\,J0658.  The value of the flux density measured by SABOCA is
  consistent with the Herschel and BLAST values.
\item Owing to the extremely high magnification, this galaxy is a
  target of choice for further studies of the properties of
  intrinsically faint high-redshift galaxies, for example with ALMA.

\end{itemize}

\begin{acknowledgements}
  We thank Anthony Gonzalez and collaborators for providing us with
  the WFC3 data displayed in Figures~\ref{fig:hstimage} and
  \ref{fig:arc}, and Padelis Papadopoulos for useful comments. We
  thank the ATCA Head of Science Operations Phil Edwards for
  allocating Director's Discretionary Time in March 2011. We thank the
  anonymous referee for his/her comments that helped to improve the
  manuscript.
\end{acknowledgements}

\bibliographystyle{bibtex/aa}
\bibliography{smmj0658}

\end{document}